\documentclass[iop]{emulateapj}

\def\kms {{\rm km~s$^{-1}$}}
\def\msun {\rm {$M_{\odot}$}}

\def\wmm {$\rm W~m^{-2}~\micron^{-1}$}

\def\oi  {[\ion{O}{1}]}

\begin{document}

\title{On the asymmetry of the OH ro-vibrational lines in HD 100546$^{\star}$}

\author{D. Fedele\altaffilmark{1}, S. Bruderer\altaffilmark{1}, M.E. van den Ancker\altaffilmark{2}, I. Pascucci\altaffilmark{3}}

\altaffiltext{$^{\star}$}{Based on observations collected at the European Southern Observatory, Paranal, Chile (Proposal ID: 075.C-0172, 084.C-0685A, 088.C-0277, 090.C-0571, 093.C-0674)}
 
\altaffiltext{1}{Max Planck Institut f\"{u}r Extraterrestrische Physik, Giessenbachstrasse 1, 85748 Garching, Germany fedele@mpe.mpg.de}
\altaffiltext{2}{European Southern Observatory, Karl Schwarzschild Strasse 2, D-85748, Garching bei M\"unchen, Germany mvandena@eso.org}
\altaffiltext{3}{Lunar and Planetary Laboratory, The University of Arizona, Tucson, AZ 85721, USA pascucci@lpl.arizona.edu}

\begin{abstract}
We present multi-epoch high-spectral resolution observations with VLT/CRIRES 
of the OH doublet $^2\Pi_{3/2}$ P4.5 (1+,1-) (2.934\,\micron) towards the 
protoplanetary disk around HD 100546. The OH doublet is detected at all 
epochs and is spectrally resolved while nearby H$_2$O lines remains 
undetected. The OH line velocity profile is different in the three datasets: 
in the first epoch (April 2012, PA=26$^{\circ}$) the OH lines are symmetric 
and line broadening is consistent with the gas being in Keplerian rotation 
around the star. No OH emission is detected within a radius of $8-11$\,au 
from the star: the line emitting region is similar in size and extent to that 
of the CO ro-vibrational lines. In the other two epochs (March 2013 and April 
2014, PA=$90^{\circ}$ and 10$^{\circ}$, respectively) the OH lines appear 
asymmetric and fainter compared to April 2012. We investigate the origin of 
these line asymmetries which were taken by previous authors as evidence for 
tidal interaction between an (unseen) massive planet and the disk. We 
show that the observed asymmetries can be fully explained by a misalignment 
of the slit of order 0\farcs04-0\farcs20 with respect to the stellar 
position. The disk is spatially resolved and the slit misalignment is likely 
caused by the extended dust emission which is brighter than the stellar photosphere 
at near-infrared wavelengths which is the wavelength used for the pointing.
This can cause the photo-center of HD 100546 to be mis-aligned with the stellar 
position at near-infrared wavelengths.
\end{abstract}


\section{Introduction}
HD 100546 is a 2.4\,\msun \ pre-main-sequence star (PMSs) surrounded by a gas- and dust-rich protoplanetary disk. The inner region
of the disk ($r \lesssim 10\,$au) is devoid of molecular gas \citep{Vanderplas09, Brittain09} while atomic gas is still present in this 
gap \citep{Acke06}. Emission by small dust grains is detected in the vicinity of the star \citep[$r < 4\,au$][]{Benisty10} while no
small grains are detected outwards up to $\sim 13\,$au from the star \citep[e.g.,][]{Bouwman03,Tatulli11,Panic14}. 
The lack of molecular gas is likely due to photo-dissociation of molecules which, given the reduced amount of small 
dust grains, are not shielded from the dissociative radiation of the star \citep[e.g.][]{Bruderer13}. 
\citet{Liskowsky12} and \citet{Brittain14} detect OH ro-vibrational emission (doublets $^2\Pi_{3/2}$ P10.5 and P9.5) 
toward this source using PHOENIX at the Gemini South telescope. The OH lines show a highly asymmetric velocity 
profile deviating from the characteristic symmetric double-peaked Keplerian profile: the blue-shifted peak is stronger than
the red-shifted one. \citet{Liskowsky12} and \citet{Brittain14} interpret the asymmetric OH profile as due to gas 
emission in a highly eccentric orbit ($e \gtrsim 0.18$) from the tidal interaction between a massive planet and the disk. 
Theoretical calculations \citep[e.g.,][]{Lubow91} and hydro-dynamical simulations \citep[e.g.][]{Kley06} show indeed that a 
stellar companion or massive planet is able to perturb the orbital motion of the gas and this produces an asymmetric velocity 
profile of the molecular gas orbiting around the star \citep{Regaly10}. Observationally, this effect was detected, e.g., 
toward the protoplanetary disk around V380 Ori \citep{Fedele11}, where the OH P4.5 line profile is asymmetric. The origin 
of this asymmetry is most likely the stellar companion of V380 Ori which perturbs the gas motion in the inner disk. 

\begin{figure*}
\centering
\includegraphics[width=18cm]{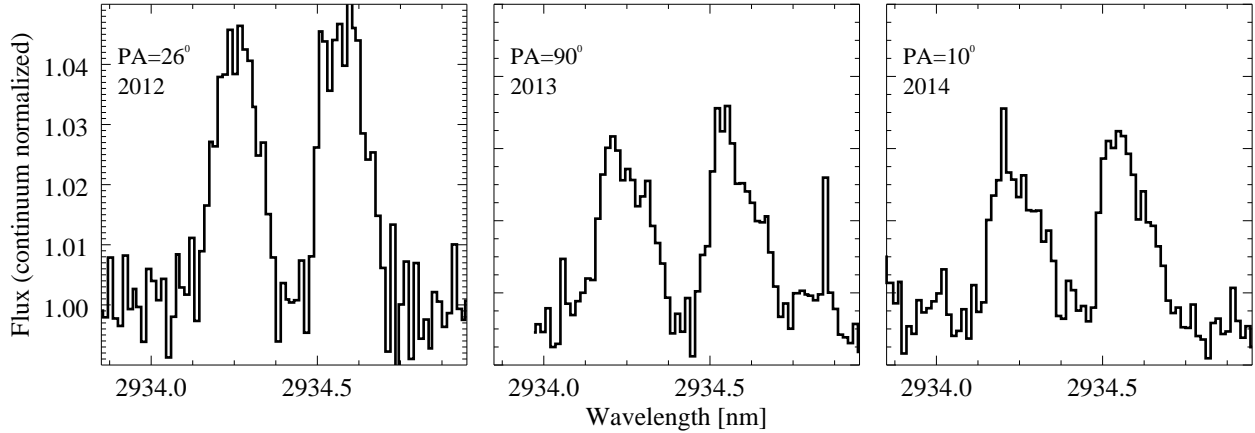}
\caption{CRIRES spectra of the OH doublet $^2\Pi_{3/2}$ P4.5 toward HD 100546 at three different position angles (and epochs).}\label{fig:oh}
\end{figure*}

\smallskip
\noindent
In the case of HD 100546 however, no stellar companion has been detected so 
far inside the dust gap \citep[e.g.][]{Grady05}, and the disk eccentricity must be caused by a massive (unseen) planetary 
companion inside the dust gap. No line asymmetry has been reported for the CO 
ro-vibrational lines (which trace similar radial distances as the OH ro-
vibrational lines) toward the  same disk by several authors 
\citep{Vanderplas09, Brittain09, Goto12,  Liskowsky12, Brittain13}. \citet{Hein14} re-observed 
HD 100546 with CRIRES and they found that the line profile of CO ro-
vibrational transitions vary (become asymmetric) with the slit position 
angle (PA) and with time (timescale of 2 nights). According to \citet{Hein14}
, this is due to a small ($\gtrsim$ 0\farcs1) offset of the slit with respect 
to the barycenter of the system; note that at a distance of 97\,pc 
\citep{Vanleeuwen07}, the inner disk of HD 100546 is spatially resolved with 
an 8-m class telescope at near-infrared wavelengths \citep[e.g.][]{Goto12}.
 
\smallskip
\noindent
This paper presents multi-epoch high-spectral resolution observations with VLT
/CRIRES of the OH ro-vibrational lines $^2\Pi_{3/2}$ P4.5 (1+,1-) at 2.934
\,\micron ~towards HD 100546. The goal is to investigate the origin of the 
asymmetry in the OH ro-vibrational lines as reported by \citet{Liskowsky12,Brittain14}.
 
\section{Observations and data reduction}\label{sec:obs}
HD 100546 was observed in the L-band at three different epochs with VLT/
CRIRES spanning two years (April 2012, March 2013, April 2014). Each 
observation of HD 100546 is followed by that of a standard star of early 
spectral type for the removal of telluric absorption features. 
The 2012 and 2014 spectra were taken with a 0\farcs2 slit width and are 
supported by the CRIRES adaptive optics system using the target as AO 
wavefront sensor and slit viewer guide star. The slit was oriented along the 
parallactic angle to minimize slit losses due to atmospheric refraction. The 
spectral resolution, measured on the OH sky emission lines is 3.5\,\kms and 
the full-width-half-maximum (FWHM) of the target is $\sim$ 0\farcs17. The 
FWHM of the telluric standard star is $\sim$ 0\farcs13, thus HD 100546 is 
spatially resolved. The 2013 spectrum (also presented in \citealt{Brittain13}) was taken with a slit width of 0\farcs4 
at a position angle of PA=$90^{\circ}$ (different from the parallactic angle) 
without adaptive optics. In this case the spectral resolution is  5.3\,\kms \ 
and the FWHM (the seeing) is 0\farcs65, thus the spectrum is affected by slit 
losses. At all epochs the centering of the target in the slit is done through 
the slit-viewer camera in the K$_s$ band. The observation log is reported in 
Table~\ref{tab:log}. 

\smallskip
\noindent
The spectra are reduced with the CRIRES data reduction pipeline using a standard
procedure: bad-pixel and cosmic rays subtraction, flat-fielding, wavelength 
calibration (using the sky emission lines as reference) and spectrum 
extraction. 
The telluric absorption lines are removed by dividing the spectrum of HD 100546
with that of the telluric standard after applying small correction to account for the
slightly different optical depth of the telluric lines between the two spectra.

\subsection{Pointing and guiding accuracy}
For CRIRES, the centering of point-sources within the slit is known to be accurate to a 
small fraction of the slit width: according to the CRIRES User 
Manual\footnote{http://www.eso.org/sci/facilities/paranal/instruments/crires/doc.html} 
the pointing and guiding accuracy is, on average of the order of 0.2 pixels (for the 0\farcs2 
slit with adaptive optics) which corresponds to 0\farcs017 at a pixel scale of 0\farcs086. We 
estimated the pointing accuracy directly from the offset applied to the telescope during the 
guiding (log files provided by J. Smoker and the ESO user support department). We find an 
accuracy of 0\farcs03 for the two AO-fed observations and 0\farcs14 for the no-AO one. Thus, 
as expected, the pointing and guiding accuracy is more accurate in the AO-supported observations. 

\subsection{Flux calibration}
The telluric standard stars are also used to calibrate the continuum flux of HD 100546
in a region free of telluric absorptions. 
The spectra extracted from the CRIRES data reduction pipeline are divided 
by the DIT of the exposure (Table~\ref{tab:log}). For each telluric standard we estimate 
the absolute spectrophotometric flux ($F_{\lambda, 0}^{kur}$) by scaling the corresponding 
Kurucz stellar atmosphere model to the 2MASS J, H and K magnitude of the star. 
The transmission function (instrument + atmosphere) $T_{\lambda}$ is given by the ratio of 
the observed (in units of ADU/s) and absolute flux (\wmm)	

\begin{equation}\label{eq:1}
{\rm T_{\lambda} = \frac{F^{obs}_{\lambda}(STD)}{F^{kur}_{\lambda, 0}(STD)}}
\end{equation}

\noindent
The science spectrum is then flux calibrated as follow:

\begin{equation}\label{eq:2}
{\rm F_{\lambda, 0}(SCI) = \frac{F_{\lambda}^{obs}(SCI)}{T_{\lambda}}}
\end{equation}
 
\noindent
where $F^{obs}_{\lambda}(SCI)$ is the observed flux after normalization for exposure time (ADU/s). 
The continuum flux of HD 100546 is then measured at $\lambda = 2934.63\,$nm, next to the OH doublet, 
in a region free on telluric absorption. Results are listed in Table~\ref{tab:phot} for the 
three epochs.

\smallskip
\noindent
The accuracy of the method was tested using the CRIRES spectra of telluric standard stars observed on 
a different night (Dec 5$^{th}$ 2008). Multiple (9) standard stars were observed during the night allowing
us to estimate the validity and precision of the flux calibration method. The spectra are presented in \citet{Fedele11}.

\noindent
Following the same procedure, the absolute flux of each star is estimated by scaling the Kurucz atmosphere model
to the J,H and K magnitudes ($F^{kur}_{\lambda, 0}(STD)$). The transmission function is measured as in eq.~\ref{eq:1} 
using one of the 9 star. The remaining 8 stars are flux calibrated using eq.~\ref{eq:2}, yielding $F_{\lambda,0}(STD)$.
For each star, we measure the flux difference at $\lambda=2934.63\,$nm as the difference: 

\begin{equation}\label{eq:3}
{\rm \Delta F(STD) = ABS\Big(\frac{F^{kur}_{0}(STD) - F_{0}(STD)}{F^{kur}_{0}(STD)}}\Big)
\end{equation}

\begin{figure}
\centering
\includegraphics[width=8cm]{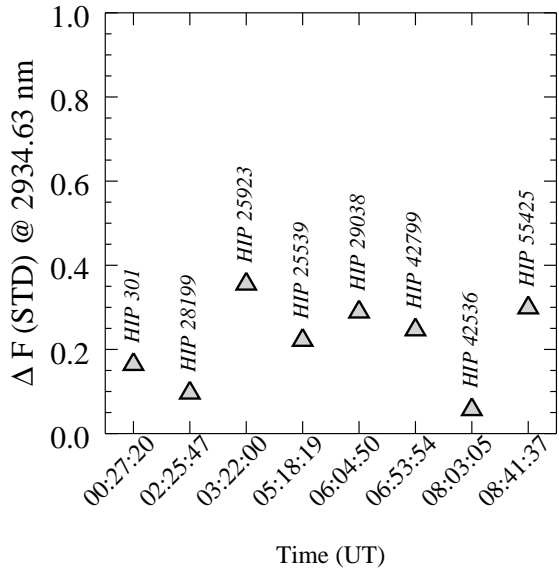}
\caption{Accuracy of the flux calibration method described in sec.~\ref{sec:obs}}
\label{fig:flux}
\end{figure}

The result is shown in Figure~\ref{fig:flux} for the different standard stars as a function of observing time (the airmass 
varies between 1.03 and 1.46). The flux difference ranges between $\sim 10-40\,$\% with an average value of 20\%. 
Thus the flux calibration method of the CRIRES spectra performed using the telluric standard stars has, on average, an 
accuracy of 20\%. 

\begin{deluxetable*}{lcccccccccc}
\centering
\tablewidth{0pt}
\tablecaption{Observations log \label{tab:log}}
\tablehead{\colhead{Date} & \colhead{Time} & \colhead{$\lambda_{\rm ref}$} &
\colhead{Slit}       & 
\colhead{Slit}       & 
\colhead{PSF}        & 
\colhead{$\Delta v$} &
\colhead{DIT}        &
\colhead{NDIT}       & 
\colhead{STD}        &
\colhead{PSF (STD)}  \\
      &
 [UT] &
 [nm] &
Width [\arcsec]      & 
PA [$^\circ$]        & 
[\arcsec]            & 
[\kms]               & 
[s] &
    &
		&
[\arcsec]} 
\startdata
2012-04-02 & 04:54:05 & 2911.5 & 0\farcs2 & 26$^{\circ}$ & 0\farcs17\tablenotemark{$\ddagger$} & 3.5 & 30 & 2 & HIP 57936 & 0.14\\
2013-03-18 & 23:38:37 & 2947.0 & 0\farcs4 & 90$^{\circ}$ & 0\farcs65                           & 5.3 & 30 & 4 & HIP 60718 & 0.63\\
2014-04-17 & 03:05:23 & 2950.0 & 0\farcs2 & 10$^{\circ}$ & 0\farcs17\tablenotemark{$\ddagger$} & 3.5 & 60 & 1 & HIP 57851 & 0.15
\enddata
\tablenotetext{$\ddagger$}{Adaptive optics supported. PA equal to the parallactic angle}
\end{deluxetable*}

\begin{deluxetable*}{lccl}
\centering
\tablecaption{Multi-epoch 3\,\micron \ flux of HD 100546 \label{tab:phot}}
\tablehead{
\colhead{Epoch} &
\colhead{F$_{\nu}$} &
\colhead{$\lambda_{ref}$} &
\colhead{Reference} \\
&
[Jy] &
[micron] &
} 
\startdata
1989-92    & 5.5             & 3.7   & \citet[][L-band photometry]{Malfait98a}   \\
1998       & 5.44 $\pm$ 0.06 & 2.934 & \citet[][ISO spectrum]{Malfait98b}\\
2010       & 6.2 $\pm$ 0.6   & 3.353 & WISE W1 photometry\\
2012       & 4.5 $\pm$ 1.0   & 2.934 & this work \\ 
2013       & 5.0 $\pm$ 1.0   & 2.934 & this work \\ 
2014       & 5.0 $\pm$ 1.0   & 2.934 & this work
\enddata
\end{deluxetable*}

\begin{deluxetable}{lcc}
\tablecaption{OH line flux for HD 100546 \label{tab:flux}}
\tablehead{
\colhead{Epoch} &
\colhead{W(OH)\tablenotemark{$\star$}}                   & 
\colhead{F(OH)}     \\
&
[$10^{-6}\,$\micron]      &
[$10^{-18}{\rm \,W\,m^{-2}}$]
} 
\startdata
2012-04-02 & $7.2 \pm 0.4$ & 11.1 $\pm$ 0.6\tablenotemark{$\dagger$}\\
2013-03-18 & $4.3 \pm 0.3$ &  7.4 $\pm$ 0.5\tablenotemark{$\dagger$}\\
2014-04-17 & $3.9 \pm 0.3$ &  6.7 $\pm$ 0.5\tablenotemark{$\dagger$}
\enddata
\tablenotetext{$\star$}{The value of equivalent width is the average of the two transitions and the error is given by the difference of the two.}
\tablenotetext{$\dagger$}{The error does not include the 20\% flux calibration uncertainty}
\end{deluxetable}

\section{Results}\label{sec:results}
\subsection{Line detection}
The OH $^2\Pi_{3/2}$ P4.5 doublet is detected in all three epochs while H$_2$O 
is undetected confirming the trend of high OH/H$_2$O abundance ratio in the 
atmosphere of the inner disk of Herbig AeBe systems \citep{Mandell08, 
Pontoppidan10, Fedele11}. Figure~\ref{fig:oh} shows the spectrum of the OH 
doublet at the three epochs. In all cases the lines are spectrally resolved. 
In April 2012 the two OH lines are symmetric contrary to what found by \citet{
Liskowsky12} in their PHOENIX spectrum (from 2010). In the other two epochs, 
the lines are asymmetric. 

\noindent
The equivalent width ($W$) is measured by integrating over the velocity range between -20\,\kms and 
+20\,\kms. The values of $W$ reported in Table~\ref{tab:flux} refer to the average value of the two 
transitions and the error is measured as the difference between the two. The line flux is measured 
multiplying $W$ by the continuum flux next to the line at $\lambda = 2934.63\,$nm. The equivalent 
width (hence the line flux) varies considerably among the three epochs with the OH line being the 
strongest in 2012 when the velocity profile is symmetric.

\subsection{Continuum emission}
The continuum flux next to the line is constant among the three epochs (Table~\ref{tab:flux}).
Multi-epoch 3\,micron flux measurements of HD 100546 are listed in Table~\ref{tab:phot}.
Our estimates are in good agreement with the ISO spectrum \citep[e.g.,][]{Malfait98b} which shows a flux 
density of 5.45\,Jy ($\pm$ 0.06\,Jy) at the same reference wavelength ($\lambda=2934.63\,$nm). The L-band 
magnitude reported by \citet{Malfait98a} is 4.15 \,mag (no error given) which corresponds to a flux density 
of 5.5\,Jy at $\lambda_e = 3.7\,$\micron \citep[adopting a zero magnitude star flux F$_0 = 253\,$Jy,][]{Lebertre98}. 
The WISE W1 magnitude is 4.2 $\pm$ 0.1\,mag \citet{Cutri12} which yields F$_{\nu} = 6.2 \pm 0.60\,$Jy (after color 
correction using the ISO spectrum) at $\lambda_e = 3.353\,$\,micron. We note that the WISE photometry  
is likely affected by the strong PAH feature between 3-3.5\,\micron \ \citep[see e.g.,][]{Malfait98b}. Indeed, 
the WISE W1 synthetic photometry computed by convolving the ISO spectrum with the W1 bandbass \citep{Wright10} gives 
a flux density of F$_{\nu}$= 5.75\,Jy, closer to the photometric measurement.

\smallskip
\noindent
In conclusion, our estimate of the continuum flux is in good agreement with the ISO spectrum and with the broad-band photometry
measured by \citep{Malfait98a} and more recently with WISE. These results do not confirm the 50\% decrease of the L-band flux 
found by \citet{Brittain13} between $\sim$ 1990 and 2010.

\section{Analysis}
The following analysis is divided in two parts: first we analyze the OH line to derive the line emitting region 
and to search for similarities/differences with respect to other gas tracers. Then, we investigate the origin of the 
asymmetric OH line profiles seen in some but not all datasets.

\begin{figure*}
\centering
\includegraphics[width=16cm]{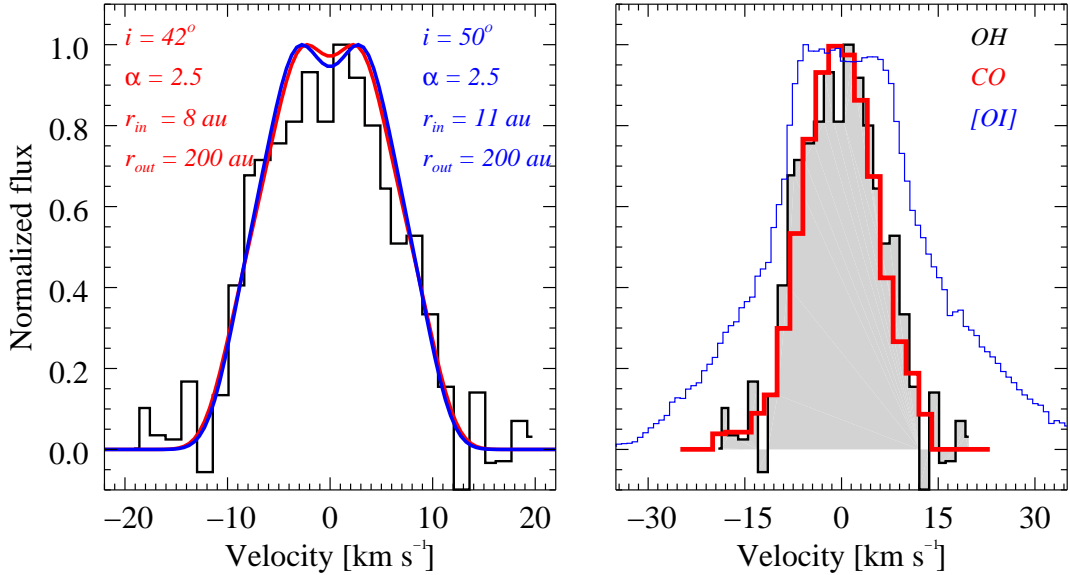}
\caption{({\it left}) OH P4.5 line velocity profile (average of the two transitions 
at PA=$26^{\circ}$) and best-fit models (solid curves) for a disk inclination of 
42$^{\circ}$ (red) and 50$^{\circ}$ (blue), respectively. ({\it right}) Comparison of the
line profiles of different gas tracers: OH (PA=$26^{\circ}$), CO (median profile of
fundamental ro-vibrational lines, PA=$55^{\circ}$) and \oi~630.0\,nm (PA=$15^{\circ}$).}
\label{fig:profile}
\end{figure*}

\subsection{OH emitting region}
To estimate the line emitting region, we create a synthetic profile assuming 
a power-law intensity profile\footnote{The formalism is described in \citet{
Fedele11}.}: 

 \begin{equation}\label{eq:intensity}
    I(r) = I(r_{\rm in}) \cdot (r/r_{\rm in})^{-\alpha}
  \end{equation}
 
\noindent
where $r$ is the distance from the star and $I(r_{\rm in})$ is the intensity 
at the inner radius.  Eq.~\ref{eq:intensity} is converted into a line 
velocity profile: assuming Keplerian rotation the projected velocity is 

\begin{equation}\label{eq:velocity}
v_{\rm proj}(r,\theta,i) = \sqrt{\frac{G M_*}{r}} \sin(i) \cos(\theta)  
\end{equation}

\noindent
with $i$ the disk inclination ($i=0^\circ$ is edge on). The OH temperature is 
fixed to 1100\,K similar to the CO temperature at the inner rim of the disk, $
\sim 10\,$au \citep[e.g.][]{Goto12, Fedele13b}. This temperature corresponds 
to a thermal broadening $v_{\rm th} = \sqrt{ 2kT/m_{\rm OH}}$ =  1.2\,\kms. 
The velocity profile is convolved with a velocity width $v = \sqrt{v_{\rm in}^
2 + v_{\rm th}^2}$ with $v_{\rm in}$ the instrumental broadening (Table~\ref{tab:log}).  

\smallskip
\noindent
To find the best fit model, the observed velocity profile is fitted by a 
synthetic profile. The parameters of the fit are: 1) the power-law index of 
the intensity $\alpha$, 2) the disk inclination, 3) the inner and 4) outer 
radius of the OH emitting region. The disk inclination and inner radius are 
partly degenerate, for this reason we fix the inclination and let the other 
parameters free. To search for the best fit parameters we create a grid of 
model line profiles varying $\alpha$ (range $1.5-4$, step 0.025), $r_{\rm in}$
 (range $5-15$\,au, step 0.1\,au) and $r_{\rm out}$ (range $30-230$\,au, step 
20\,au) and we compute the reduced $\chi^2$ between the observed and model 
line profile. The set of parameters that best fit the observed profile 
are found minimizing $\tilde{\chi}^2$

\begin{equation}
\tilde{\chi}^2 = \frac{1}{N-1-n}\Sigma_i \Big(\frac{m_i - f_i}{\sigma_i}\Big)^2
\end{equation}

with $N$ the number of velocity bins in the range between $-25$\kms and $+25$
\kms, $n$ the number of parameters ($=4$), $m_i$, $f_i$ the model predicted 
and observed flux at velocity bin $i$ and $\sigma_i$ the corresponding 
uncertainty. The fitting procedure is repeated twice for two different 
values of the inclination, 42$^{\circ}$ and 50$^{\circ}$, since all inclinations
reported in the literature are within this range \citep{Pantin00, Augereau01, Grady01, Liu03, 
Ardila07, Panic14, Avenhaus14}. The parameters that best fit the observed profile are 
listed in Figure~\ref{fig:profile} together with the best fit models: for 
the $i=42^{\circ}$ case we find an inner radius of 8\,au ($\tilde{\chi}^2 = 1.
02$), while for the $i=50^{\circ}$ case we find an inner radius of 11\,au ($
\tilde{\chi}^2 = 1.03$). 

\smallskip
\noindent
The lack of high-velocity gas indicates a gap (drop in abundance) of OH gas 
inside a gap of radius $\sim 8-11$\,au (depending on the assumed inclination).

\subsection{Comparison of CRIRES and PHOENIX OH profile}
\begin{figure}
\centering
\includegraphics[width=8cm]{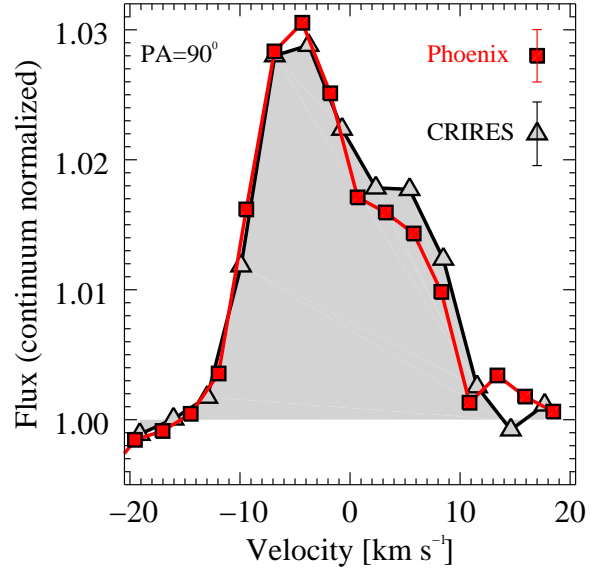}
\caption{Comparison of CRIRES and PHOENIX spectra at PA=90$^{\circ}$. 
The average of the two P4.5 transitions (1+, 1-) is shown for the CRIRES 
spectra while the PHOENIX spectrum is the average of the four lines detected 
by \citet{Liskowsky12} (doublet P9.5 and P10.5). In all the spectra the 
continuum is normalized to unity.}
\label{fig:phoenix}
\end{figure}

Figure \ref{fig:phoenix} shows a direct comparison of the CRIRES and PHOENIX 
spectra (presented in \citealt{Liskowsky12}) taken with the same slit 
position angle (90$^{\circ}$ and width (0\farcs4): to increase the S/N we 
averaged the two lines of the OH P4.5 doublet (CRIRES), while the PHOENIX 
spectrum is the average of the four lines detected by Liskowsky et al. 
(doublets P9.5 and P10.5). Finally, the spectra are binned in wavelength to 
further increase the S/N. The two spectra are similar. There are however some 
differences: the peak-to-peak asymmetry appears more pronounced in the 
PHOENIX spectrum, where the red-shifted component is systematically fainter 
than in the CRIRES spectra. This difference is significant as it is seen in 
several (consecutive) spectral bins. 

\subsection{Comparison to other gas tracers}
In this section we compare the velocity profile of the OH P4.5 doublet to 
that of other gas tracers, namely the CO ro-vibrational lines and the optical 
forbidden line [\ion{O}{1}]\,630.0\,nm. The median profile of the CO 
fundamental ro-vibrational transitions (v=1-0, 2-1, 3-2 and 4-3) is shown in 
Figure~\ref{fig:profile} (right). These spectra were taken with CRIRES with a 
spectral resolution of $\sim$ 3\,\kms \ (slit width = 0\farcs2, PA=$55^{\circ
}$, PSF=0\farcs17, March 29$^{th}$ 2010, \citealt{Hein14}). The similarity of 
the velocity profiles of the OH (PA=$26^{\circ}$) and CO ro-vibrational 
lines and the symmetric profile of CO suggests that these transitions come 
from a similar radial extent in the disk. The inner radius of the OH emitting region 
found here is in good agreement with $r_{\rm in}({\rm CO)}$ from previous 
estimates: $r_{\rm in}(\rm CO) = 8\,au$ \citep[][$i=42$]{Vanderplas09}, $r_{
\rm in}(\rm CO) = 13\,au$ \citep[][$i=50$]{Brittain09}. 

\smallskip
\noindent
The spectrum of the \oi\,630.0\,nm line is also shown in Figure~\ref{fig:profile} 
(VLT/UVES, PA = 15$^{\circ}$, slit width = 0\farcs3, \citealt{Acke06}). The 
oxygen line extends to high velocity indicating the presence of atomic gas 
inside the disk gap. The low-velocity part of the \oi \ line appears slightly 
asymmetric and both the asymmetry and the line intensity vary with 
time according to \citet{Acke06}. The asymmetry in the \oi \ line is however 
much less pronounced than that of the OH P4.5 lines at slit PA = $10^{\circ}$ 
and $90^{\circ}$. Moreover, the temporal changes reported by \citet{Acke06} 
are not reconcilable with the changes in the OH line as the OH asymmetry is 
only observed in the blue-shifted peak, contrary to the \oi \ where 
the asymmetry is observed also in the red-shifted peak. For further 
discussion see Sec.~\ref{sec:discussion}. 

\subsection{Line asymmetry}\label{sec:offset}
As shown by \citet{Hein14}, in the case of spatially resolved observations, a 
slight offset can induce slit losses in the CO ro-vibrational lines and 
produce an asymmetric line profile. Given the similar size and extent, also 
the OH ro-vibrational lines may be affected by the same process. 
The different flux of the OH line between the three epochs 
can be the consequence of slit losses. If this is the case the 2012 spectrum, 
showing the higher line flux, is the less affected one. 
To investigate this, we create a synthetic disk image using a geometrical model 
where the OH line is assumed to emerge from either the surface or an inner 
wall of a disk in Keplerian rotation. The intensity on the surface and the 
wall are taken to be the same. Figure~\ref{fig:disk} shows the synthetic 
velocity map. The figure also shows the width and orientation of the three 
CRIRES OH spectra. The synthetic disk image is convolved with a 2-D Gaussian 
profile to mimic the point-spread-function (PSF) of the CRIRES spectra. The 
FWHM of the convolution is given by the actual size 
of the target in the spatial direction of the spectrum (Table~\ref{tab:log}): 
this corresponds to 0\farcs17 for the spectra taken with the 0\farcs2 slit (
PA = 26$^{\circ}$ and 10$^{\circ}$) and 0\farcs65 in the case of the 0\farcs4 
slit (PA = 90$^{\circ}$). The synthetic line profiles are created by 
filtering the convolved disk images with the corresponding slit width and 
position angle. The effect of a slit misalignment, is reproduced by an offset 
between the slit and the stellar position, in the direction perpendicular to 
the slit PA. The sign convention for the offset is the same as in \citet{Hein14}. 
Finally, the synthetic line profiles are convolved with the corresponding spectral 
resolution: 3.5\,\kms \ and 5.3\,\kms \ for the 0\farcs2 and 0\farcs4 slit width, 
respectively (Table~\ref{tab:log}). 

\begin{figure*}
\centering
\includegraphics[width=16cm]{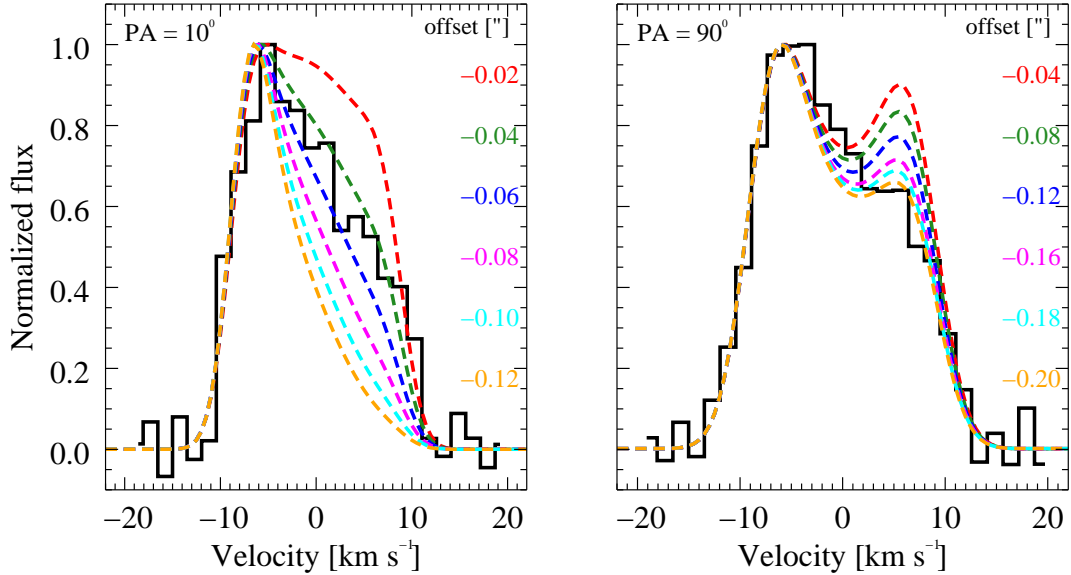}
\caption{Comparison of the PA=$10^{\circ}$ (left) and PA=90$^{\circ}$ (right) 
CRIRES OH spectra (average profile of the 1+ and 1- transitions) with 
synthetic profiles with different offset between the slit position and the 
central star (see Sec.~\ref{sec:offset}). }\label{fig:offset}
\end{figure*}

\smallskip
\noindent
Figure~\ref{fig:offset} shows the synthetic spectra for different offsets in 
the cases of PA=10$^{\circ}$ and PA=90$^{\circ}$: the offset produces an 
asymmetry in the line profile, and the flux ratio between the blue- and red-
shifted component of the OH line increases with the offset. 
For a given offset, the asymmetry is more pronounced in the narrow slit spectrum. 
This is due to the different size of the PSF.
The spectra are normalized although the offset induces slit losses which results in a 
fainter line flux as the offset increases. A small offset (between -0\farcs04 
and -0\farcs06) is enough to reproduce the observed profile at PA=10$^{\circ}$, 
while a larger offset (between -0\farcs16 and -0\farcs20) is needed for the 
PA=$90^{\circ}$ spectrum.

\section{Discussion}\label{sec:discussion}
The analysis presented here demonstrates that the asymmetric profile of the 
OH ro-vibrational lines are consistent with an offset between the slit and 
the stellar position. Our interpretation is that the spectra which are mostly 
affected by the slit mis-alignment are the PA=10$^{\circ}$ and PA=90$^{\circ}$
 (both CRIRES and PHOENIX). The PA=26$^{\circ}$ spectrum shows indeed no 
clear evidence of asymmetry in the line profile which appears instead top-
flat as we would expect in the case of Keplerian rotation. Moreover, in the 
same spectrum the line is much stronger (almost double, Table~\ref{tab:log}) 
than in the other spectra. This finding suggest that the PA=10$^{\circ}$ and 
PA=90$^{\circ}$ spectra can be affected by slit losses which is a natural 
consequence of the slit mis-alignment. 
The small offsets estimated in the previous section produce negligible or no 
change of the continuum flux but considerable variation of the OH line velocity 
profile. 
  
\subsection{Origin of the OH line asymmetry}
\begin{figure}
\centering
\includegraphics[width=9cm]{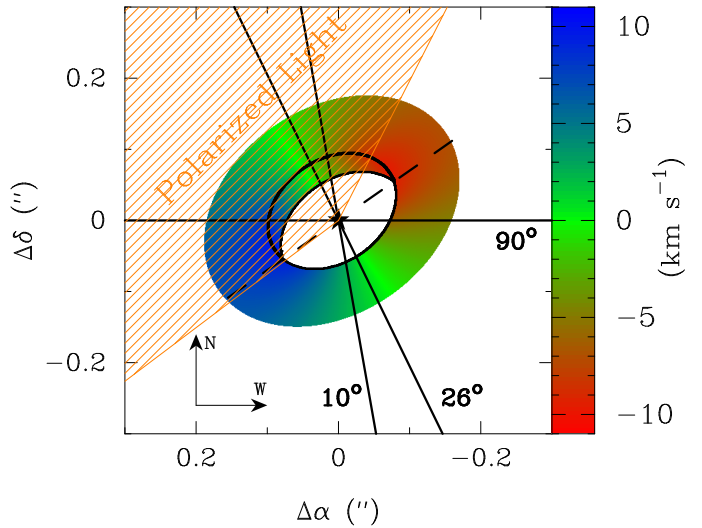}
\caption{Synthetic OH velocity map. The dashed line shows the position 
angle of the disk major disk \citep[145$^{\circ}$, ][]{Ardila07}. The slit 
position angles at the three epochs are also shown. The dashed region 
shows the side of the disk which appears brighter in polarized light from
\citet{Quanz11} and \citet{Avenhaus14}.}
\label{fig:disk}
\end{figure}

If the asymmetric profile of the OH line is due to an offset of the slit, 
this must be caused by the source itself as it is observed with two different 
instruments (CRIRES and PHOENIX). 
A possible explanation is that the misalignment of the slit is 
caused by a non homogenous illumination: we infer that this is due to the 
finite size of the disk inner wall at $\sim 10$\,au from the star. 
Because of the disk inclination ($\sim 42^{\circ} - 50^{\circ}$) the inner 
wall at 10-14\,au produces an asymmetric image in the sky (see 
Figure~\ref{fig:disk}). Interestingly, deviation from axis-symmetric emission is detected 
in mid-infrared interferometric observations \citep{Panic14}. Given the 
higher intensity of the disk wall over the stellar photosphere at wavelength 
$> 1\,$\micron, the peak of the continuum intensity at these wavelengths does 
not coincide with the stellar position. Since the telescope pointing is 
automatically adjusted toward the peak of the emission in the K$_{\rm s}$
band in the slit viewer camera\footnote{Note that the star itself is occulted 
by the slit in the SV camera.}, this can affect the centering of the target 
inside the slit, inducing an offset between the center of the slit and the 
stellar position. The misalignment of the slit should be maximum when the slit PA is 
parallel to the disk major axis \citep[$\sim 130^{\circ}-160^{\circ}$, e.g.,
][]{Pantin00, Augereau01, Grady01, Ardila07, Panic14, Avenhaus14} and it 
should be minimized when the slit is oriented along the disk minor axis. This 
is consistent with the different offset measured in the PA=$90^{\circ}$ and PA
=$10^{\circ}$ spectra and with the symmetric profile seen in the PA=$26^{\circ
}$ (closer to the disk minor axis) spectrum.

\smallskip
\noindent
Our interpretation is supported by the polarimetric differential imaging (PDI)
 of HD 100546 performed by \citet{Quanz11} and \citet{Avenhaus14}: the authors detect a 
brightness asymmetry in polarized light with the North-East part (far side) 
of the disk being brighter than the Southern-West one (near side). The 
azimuthal directions of this asymmetry is shown in Fig.~\ref{fig:disk}.  

\smallskip
\noindent
We further note that our interpretation is in good agreement with near-infrared 
interferometric observations \citep{Benisty10, Tatulli11}: the K-band VLTI/
AMBER visibility at short baselines can only be fitted by a model which 
includes spatially extended emission, consistent with the disk inner wall at $
\sim 13\,$au \citep{Tatulli11}.

\subsection{Disk wall or clumps}
There are	 other possible scenarios that can induce a slit offset: 
any bright source (brighter than the star in the K-band) within $\sim$ 0\farcs
2 from the star could induce a mis-alignment of the slit during the 
acquisition of the spectrum. One possibility is a stellar companion, there 
are however no detections of any bright companion within 0\farcs2 to date. 
Another possibility is inhomogeneous dust continuum emission from the disk. Recent 
polarimetric observations by \citet{Avenhaus14} rule out the presence of a disk 
hole at PA=$12^{\circ}$ claimed by \citet{Quanz11} while they detect two bright spots 
at PA$\sim -30^{\circ}$ and PA$\sim 130^{\circ}$ with the latter being brighter than 
the first one. We warn however that the spots seen in polarized light do not necessarily 
imply the existence of bright spots in the total (unpolarized) K-band flux. The results 
of \citet{Quanz11} and \citet{Avenhaus14} are instead consistent with the overall geometry 
of the disk where the North-East side (brighter in scattered light) is facing toward us. 

\subsection{Disk eccentricity}
\citet{Liskowsky12} suggest that the inner disk of HD 100546 is eccentric and 
the OH distribution is not homogeneous: the authors propose a scenario in 
which a massive (unseen) planet perturbs the gas dynamics at the distance of 
the disk inner wall. \citet{Brittain14} further support this scenario 
based on the non-variability of the OH line profiles: the OH spectra 
presented by \citet{Brittain14} are both taken with the same slit PA 
(=90$^{\circ}$). As shown in Figure~\ref{fig:phoenix} however, the PA=$90^{\circ}$ 
spectra (CRIRES and PHOENIX) do show a slightly different line profile, which cannot 
be due to disk eccentricity. Moreover, the PA=10$^{\circ}$ and PA=26$^{\circ}$ spectra show 
drastic variation in line profile and equivalent width.  

\smallskip
\noindent
\citet{Brittain14} assume an eccentricity of 0.18 ($\pm$0.11) to 
explain the asymmetry of the OH lines. However, no evidence for disk eccentricity
was found in the differential polarimetric observations of \citet[][$e < 0.133$ at 99.8\% confidence]{Avenhaus14}
  
\section{Conclusion}
Based on the data collected here and on the performed analysis, the asymmetric 
profile of the OH ro-vibrational lines toward HD 100546 are consistent with a 
misalignment of the slit of the order of $0\farcs04-0\farcs 2$. We argue 
that the misalignment results from the finite size of the disk inner 
wall at $\sim10 - 14$\,au from the star ($\sim 0\farcs10 - 0\farcs14$ at a distance 
of 97\,pc).  
Thus there is no need to invoke a highly eccentric gas disk, as was done by \citet{Liskowsky12, Brittain14}, 
to explain the asymmetric line profile in HD 100546.
 The analysis presented in this paper, however, does not exclude the presence of a massive 
planet/companion inside the disk gap as suggested by several authors \citep[e.g.,][]{Bouwman03, Acke06, Mulders13}.

{\it Facilities:} \facility{VLT/CRIRES}
\acknowledgments
We are grateful to the VLT telescope operators and astronomers who performed the CRIRES observations in service mode.
We thank R. Hein Bertelsen for providing the median CO ro-vibrational profiles an G. van der Plas for providing the \oi \ data. 
DF thanks S. Brittain for an interesting discussion and for providing the PHOENIX spectrum. We thank J. Smoker and A. Smette for 
discussion on slit centering with CRIRES and the ESO USD for providing the log files. DF thanks J. Bouwmann for providing the ISO 
spectrum of HD 100546 and T. M\"uller for further useful discussion on the WISE photometry. We are grateful to the anonymous 
referee for providing useful comments and suggestions. IP acknowledges support from a NSF Astronomy \& Astrophysics Research Grant (ID:1312962).


\begin{thebibliography}{35}
\expandafter\ifx\csname natexlab\endcsname\relax\def\natexlab#1{#1}\fi

\bibitem[{{Acke} \& {van den Ancker}(2006)}]{Acke06}
{Acke}, B., \& {van den Ancker}, M.~E. 2006, \aap, 449, 267

\bibitem[{{Ardila} {et~al.}(2007){Ardila}, {Golimowski}, {Krist}, {Clampin},
  {Ford}, \& {Illingworth}}]{Ardila07}
{Ardila}, D.~R., {Golimowski}, D.~A., {Krist}, J.~E., {Clampin}, M., {Ford},
  H.~C., \& {Illingworth}, G.~D. 2007, \apj, 665, 512

\bibitem[{{Augereau} {et~al.}(2001){Augereau}, {Lagrange}, {Mouillet}, \&
  {M{\'e}nard}}]{Augereau01}
{Augereau}, J.~C., {Lagrange}, A.~M., {Mouillet}, D., \& {M{\'e}nard}, F. 2001,
  \aap, 365, 78

\bibitem[{{Avenhaus} {et~al.}(2014){Avenhaus}, {Quanz}, {Meyer}, {Brittain},
  {Carr}, \& {Najita}}]{Avenhaus14}
{Avenhaus}, H., {Quanz}, S.~P., {Meyer}, M.~R., {Brittain}, S.~D., {Carr},
  J.~S., \& {Najita}, J.~R. 2014, \apj, 790, 56

\bibitem[{{Benisty} {et~al.}(2010){Benisty}, {Tatulli}, {M{\'e}nard}, \&
  {Swain}}]{Benisty10}
{Benisty}, M., {Tatulli}, E., {M{\'e}nard}, F., \& {Swain}, M.~R. 2010, \aap,
  511, A75

\bibitem[{{Bouwman} {et~al.}(2003){Bouwman}, {de Koter}, {Dominik}, \&
  {Waters}}]{Bouwman03}
{Bouwman}, J., {de Koter}, A., {Dominik}, C., \& {Waters}, L.~B.~F.~M. 2003,
  \aap, 401, 577

\bibitem[{{Brittain} {et~al.}(2014){Brittain}, {Carr}, {Najita}, {Quanz}, \&
  {Meyer}}]{Brittain14}
{Brittain}, S.~D., {Carr}, J.~S., {Najita}, J.~R., {Quanz}, S.~P., \& {Meyer},
  M.~R. 2014, \apj, 791, 136

\bibitem[{{Brittain} {et~al.}(2009){Brittain}, {Najita}, \&
  {Carr}}]{Brittain09}
{Brittain}, S.~D., {Najita}, J.~R., \& {Carr}, J.~S. 2009, \apj, 702, 85

\bibitem[{{Brittain} {et~al.}(2013){Brittain}, {Najita}, {Carr}, {Liskowsky},
  {Troutman}, \& {Doppmann}}]{Brittain13}
{Brittain}, S.~D., {Najita}, J.~R., {Carr}, J.~S., {Liskowsky}, J., {Troutman},
  M.~R., \& {Doppmann}, G.~W. 2013, \apj, 767, 159

\bibitem[{{Bruderer}(2013)}]{Bruderer13}
{Bruderer}, S. 2013, \aap, 559, A46

\bibitem[{{Cutri} \& {et al.}(2012)}]{Cutri12}
{Cutri}, R.~M., \& {et al.} 2012, VizieR Online Data Catalog, 2311, 0

\bibitem[{{Fedele} {et~al.}(2013){Fedele}, {Bruderer}, {van Dishoeck},
  {Hogerheijde}, {Panic}, {Brown}, \& {Henning}}]{Fedele13b}
{Fedele}, D., {Bruderer}, S., {van Dishoeck}, E.~F., {Hogerheijde}, M.~R.,
  {Panic}, O., {Brown}, J.~M., \& {Henning}, T. 2013, \apjl, 776, L3

\bibitem[{{Fedele} {et~al.}(2011){Fedele}, {Pascucci}, {Brittain}, {Kamp},
  {Woitke}, {Williams}, {Dent}, \& {Thi}}]{Fedele11}
{Fedele}, D., {Pascucci}, I., {Brittain}, S., {Kamp}, I., {Woitke}, P.,
  {Williams}, J.~P., {Dent}, W.~R.~F., \& {Thi}, W.-F. 2011, \apj, 732, 106

\bibitem[{{Goto} {et~al.}(2012){Goto}, {van der Plas}, {van den Ancker},
  {Dullemond}, {Carmona}, {Henning}, {Meeus}, {Linz}, \& {Stecklum}}]{Goto12}
{Goto}, M., {et~al.} 2012, \aap, 539, A81

\bibitem[{{Grady} {et~al.}(2005){Grady}, {Woodgate}, {Heap}, {Bowers}, {Nuth},
  {Herczeg}, \& {Hill}}]{Grady05}
{Grady}, C.~A., {Woodgate}, B., {Heap}, S.~R., {Bowers}, C., {Nuth}, III,
  J.~A., {Herczeg}, G.~J., \& {Hill}, H.~G.~M. 2005, \apj, 620, 470

\bibitem[{{Grady} {et~al.}(2001){Grady}, {Polomski}, {Henning}, {Stecklum},
  {Woodgate}, {Telesco}, {Pi{\~n}a}, {Gull}, {Boggess}, {Bowers}, {Bruhweiler},
  {Clampin}, {Danks}, {Green}, {Heap}, {Hutchings}, {Jenkins}, {Joseph},
  {Kaiser}, {Kimble}, {Kraemer}, {Lindler}, {Linsky}, {Maran}, {Moos}, {Plait},
  {Roesler}, {Timothy}, \& {Weistrop}}]{Grady01}
{Grady}, C.~A., {et~al.} 2001, \aj, 122, 3396

\bibitem[{{Hein Bertelsen} {et~al.}(2014){Hein Bertelsen}, {Kamp}, {Goto}, {van
  der Plas}, {Thi}, {Waters}, {van den Ancker}, \& {Woitke}}]{Hein14}
{Hein Bertelsen}, R.~P., {Kamp}, I., {Goto}, M., {van der Plas}, G., {Thi},
  W.-F., {Waters}, L.~B.~F.~M., {van den Ancker}, M.~E., \& {Woitke}, P. 2014,
  \aap, 561, A102

\bibitem[{{Kley} \& {Dirksen}(2006)}]{Kley06}
{Kley}, W., \& {Dirksen}, G. 2006, \aap, 447, 369

\bibitem[{{Le Bertre} \& {Winters}(1998)}]{Lebertre98}
{Le Bertre}, T., \& {Winters}, J.~M. 1998, \aap, 334, 173

\bibitem[{{Liskowsky} {et~al.}(2012){Liskowsky}, {Brittain}, {Najita}, {Carr},
  {Doppmann}, \& {Troutman}}]{Liskowsky12}
{Liskowsky}, J.~P., {Brittain}, S.~D., {Najita}, J.~R., {Carr}, J.~S.,
  {Doppmann}, G.~W., \& {Troutman}, M.~R. 2012, \apj, 760, 153

\bibitem[{{Liu} {et~al.}(2003){Liu}, {Hinz}, {Meyer}, {Mamajek}, {Hoffmann}, \&
  {Hora}}]{Liu03}
{Liu}, W.~M., {Hinz}, P.~M., {Meyer}, M.~R., {Mamajek}, E.~E., {Hoffmann},
  W.~F., \& {Hora}, J.~L. 2003, \apjl, 598, L111

\bibitem[{{Lubow}(1991)}]{Lubow91}
{Lubow}, S.~H. 1991, \apj, 381, 259

\bibitem[{{Malfait} {et~al.}(1998{\natexlab{a}}){Malfait}, {Bogaert}, \&
  {Waelkens}}]{Malfait98a}
{Malfait}, K., {Bogaert}, E., \& {Waelkens}, C. 1998{\natexlab{a}}, \aap, 331,
  211

\bibitem[{{Malfait} {et~al.}(1998{\natexlab{b}}){Malfait}, {Waelkens},
  {Waters}, {Vandenbussche}, {Huygen}, \& {de Graauw}}]{Malfait98b}
{Malfait}, K., {Waelkens}, C., {Waters}, L.~B.~F.~M., {Vandenbussche}, B.,
  {Huygen}, E., \& {de Graauw}, M.~S. 1998{\natexlab{b}}, \aap, 332, L25

\bibitem[{{Mandell} {et~al.}(2008){Mandell}, {Mumma}, {Blake}, {Bonev},
  {Villanueva}, \& {Salyk}}]{Mandell08}
{Mandell}, A.~M., {Mumma}, M.~J., {Blake}, G.~A., {Bonev}, B.~P., {Villanueva},
  G.~L., \& {Salyk}, C. 2008, \apjl, 681, L25

\bibitem[{{Mulders} {et~al.}(2013){Mulders}, {Paardekooper}, {Pani{\'c}},
  {Dominik}, {van Boekel}, \& {Ratzka}}]{Mulders13}
{Mulders}, G.~D., {Paardekooper}, S.-J., {Pani{\'c}}, O., {Dominik}, C., {van
  Boekel}, R., \& {Ratzka}, T. 2013, \aap, 557, A68

\bibitem[{{Pani{\'c}} {et~al.}(2014){Pani{\'c}}, {Ratzka}, {Mulders},
  {Dominik}, {van Boekel}, {Henning}, {Jaffe}, \& {Min}}]{Panic14}
{Pani{\'c}}, O., {Ratzka}, T., {Mulders}, G.~D., {Dominik}, C., {van Boekel},
  R., {Henning}, T., {Jaffe}, W., \& {Min}, M. 2014, \aap, 562, A101

\bibitem[{{Pantin} {et~al.}(2000){Pantin}, {Waelkens}, \& {Lagage}}]{Pantin00}
{Pantin}, E., {Waelkens}, C., \& {Lagage}, P.~O. 2000, \aap, 361, L9

\bibitem[{{Pontoppidan} {et~al.}(2010){Pontoppidan}, {Salyk}, {Blake},
  {Meijerink}, {Carr}, \& {Najita}}]{Pontoppidan10}
{Pontoppidan}, K.~M., {Salyk}, C., {Blake}, G.~A., {Meijerink}, R., {Carr},
  J.~S., \& {Najita}, J. 2010, \apj, 720, 887

\bibitem[{{Quanz} {et~al.}(2011){Quanz}, {Schmid}, {Geissler}, {Meyer},
  {Henning}, {Brandner}, \& {Wolf}}]{Quanz11}
{Quanz}, S.~P., {Schmid}, H.~M., {Geissler}, K., {Meyer}, M.~R., {Henning}, T.,
  {Brandner}, W., \& {Wolf}, S. 2011, \apj, 738, 23

\bibitem[{{Reg{\'a}ly} {et~al.}(2010){Reg{\'a}ly}, {S{\'a}ndor}, {Dullemond},
  \& {van Boekel}}]{Regaly10}
{Reg{\'a}ly}, Z., {S{\'a}ndor}, Z., {Dullemond}, C.~P., \& {van Boekel}, R.
  2010, \aap, 523, A69

\bibitem[{{Tatulli} {et~al.}(2011){Tatulli}, {Benisty}, {M{\'e}nard},
  {Varni{\`e}re}, {Martin-Za{\"i}di}, {Thi}, {Pinte}, {Massi}, {Weigelt},
  {Hofmann}, \& {Petrov}}]{Tatulli11}
{Tatulli}, E., {et~al.} 2011, \aap, 531, A1

\bibitem[{{van der Plas} {et~al.}(2009){van der Plas}, {van den Ancker},
  {Acke}, {Carmona}, {Dominik}, {Fedele}, \& {Waters}}]{Vanderplas09}
{van der Plas}, G., {van den Ancker}, M.~E., {Acke}, B., {Carmona}, A.,
  {Dominik}, C., {Fedele}, D., \& {Waters}, L.~B.~F.~M. 2009, \aap, 500, 1137

\bibitem[{{van Leeuwen}(2007)}]{Vanleeuwen07}
{van Leeuwen}, F. 2007, \aap, 474, 653

\bibitem[{{Wright} {et~al.}(2010){Wright}, {Eisenhardt}, {Mainzer}, {Ressler},
  {Cutri}, {Jarrett}, {Kirkpatrick}, {Padgett}, {McMillan}, {Skrutskie},
  {Stanford}, {Cohen}, {Walker}, {Mather}, {Leisawitz}, {Gautier}, {McLean},
  {Benford}, {Lonsdale}, {Blain}, {Mendez}, {Irace}, {Duval}, {Liu}, {Royer},
  {Heinrichsen}, {Howard}, {Shannon}, {Kendall}, {Walsh}, {Larsen}, {Cardon},
  {Schick}, {Schwalm}, {Abid}, {Fabinsky}, {Naes}, \& {Tsai}}]{Wright10}
{Wright}, E.~L., {et~al.} 2010, \aj, 140, 1868

\end{thebibliography}

\end{document}